Electrical Spin Injection into Silicon using MgO Tunnel Barrier

Tomoyuki Sasaki[*], Tohru Oikawa, Toshio. Suzuki[1], Masashi Shiraishi[2]

Yoshishige Suzuki[2], and Katsumichi Tagami

SQ Research Center, TDK Corporation, 385-8555 Nagano, Japan

[1]AIT, Akita Research Institute of Advanced Technology, 010-1623 Akita, Japan

[2]Graduate School of Engineering Science, Osaka University, 560-8531 Osaka, Japan

Abstract

We observed spin injection into silicon through Fe/MgO tunnel barrier by using non-local magnetoresistance measurement technique. Fe/MgO tunnel barrier contacts with a lateral spin valve structure were fabricated on phosphorus doped silicon-on-insulator substrate. Spin injection signals in the non-local scheme were observed up to 120 K, which is the highest value where band transferred spins in Si have ever been reported, and spin diffusion length was estimated to be about 2.25 $\mu$m at 8 K. Temperature dependence and injection current dependence of the non-local voltage were also investigated. It is clarified that MgO tunnel barrier is effective for the spin injection into silicon.

*E-mail address: tomosasa@jp.tdk.com

Spintronics is a new category in science as a fusion domain of magnetism and electronics in which a spin degree of freedom plays a key role in an electronic device. For further progress in spintronics, semiconductor spin devices using GaAs and Si have attracted much attention in the recent decade, because they allow us to fabricate novel spin devices, which can give a large impact to the commercial market based on inorganic semiconductor devices [1, 2, 3, 4]. Si is especially regarded as promising material, because its lattice symmetry is higher than that of GaAs, which induces comparably smaller spin-orbit coupling. Several groups have proposed and demonstrated Si-based spintronics devices, such as a spin metal-oxide-semiconductor field effect transistor (spin-MOSFET) [5, 6, 7], a lateral spin valve [8] and a hot-electron spin transistor [9, 10]. Although Huang *et al*. observed excellent spin coherence of hot-electron over the distance 350 μm by observing a Hanle-type spin precession effect [10], it should be noted that the observed good spin coherence was realized only in a semi-insulating pure Si wafer. Van t'Erve *et al.* have demonstrated, for the first time, spin injection, creation of a spin current and diffusive band transport of the spins in Si by using a Fe/AlO$_x$ tunnel barrier and a "non-local" (NL) measurement technique [11]. However, the band transport of spins in their study was realized only below 10 K. In addition, whilst Hanle-type spin precession was observed, precise measurements of spin diffusion length was not implemented, which was attributed to a difficulty in realization of anti-parallel magnetization alignments due to a large sample size.

In this manuscript, we report on achievements of spin injection and band transport of the injected spins up to 120 K, which is the highest temperature concerning the band transport of the injected spins to our best knowledge, by using Fe/MgO/Si tunnel contacts with a lateral spin valve structure and the "non-local" method [12]. We also report on a precise estimation of the spin diffusion length in Si, and furthermore, the temperature and bias current dependence of non-local output voltage is discussed.

Figure 1 (a) shows a schematic image of a four-terminal lateral spin valve device, and Fig.1 (b) shows an optical microscopic image of the device. The silicon spin valve transport samples were fabricated on a silicon-on-insulator (SOI) substrate with (100)-plane, of which structure was silicon(100 nm)/silicon oxide(200 nm)/Si substrate. The SOI was annealed at 900 ℃ after the P was implanted into Si, and an electron concentration of P-doped Si was approximately $1 \times 10^{20}$ cm$^{-3}$ at room temperature (RT). The temperature dependence of resistivity exhibited metallic behavior above 150 K and semiconductive behavior below 150 K. The Si surface was washed by using dilute HF solution, and rinsed in a de-ionized water and isopropyl alcohol in order to remove the thin silicon oxide film and some kinds of dust. Then a Ti(5 nm)/Fe(13 nm)/MgO(0.8 nm in Sample A and 1.4 nm in Sample B) film was deposited by using a molecular beam epitaxy (MBE) system. We confirmed that SiO$_x$ and/or oxide impurities did not exist at MgO/Si interface by the transmission electron microscopy (TEM) and energy dispersive X-ray spectroscopy (EDS). Two ferromagnetic (FM) electrodes (contact 2 and 3) were formed by ion milling. The FM electrode have different widths in order to obtain different coercive forces, and the area of them were 0.5×21 μm (Contact 2) and 2.5×21 μm (Contact 3), respectively. A Si channel was formed by a mesa-etching technique (see Fig. 1). In order to obtain electrical isolation of the Si channel, the surface of the channel was oxidized, and 30 nm-thick SiO$_2$ was deposited onto the surface. Contacts 1 and 4 are nonmagnetic (NM) electrodes (21×21 μm in size), which were placed about 150 μm away from the spin injection contact 2 and the spin detection contact 3, respectively. The pad electrodes on contact 1 to 4 were Au(150 nm)/Cr(50 nm). The interconnection between pad electrodes was formed as Ta(10 nm)/Cu(20 nm)/Ta(20 nm) wires covering all over the magnetic electrodes. Because the resistivity of Cu is smaller than that of Ta, the spin injection current is perpendicular to the FM contacts. Consequently, this device structure allows us to avoid an observation of parasitic signal caused from anisotropic magnetoresistance (AMR) effect. To measure magnetic switching of the electrodes using the AMR effect, we made separate samples without Cu/Ta layer (dummy sample).

NL measurements were performed by using a four-terminal probing system (ST-500, JANIS) with an electromagnet between 4.2 K and RT. An advantage of introducing the NL method is that one can exclude spurious signals, such as signals induced by AMR effect, and can obtain reliable results since the charge current path and the spin current path are completely separated [13, 14]. The magnetic field was swept from -80 mT to 80 mT in 0.8 mT steps. We employed a standard AC lock-in technique (lock-in frequency = 333 Hz, time constant = 300 ms). The spin polarized electrons were injected into the Si by applying a voltage between contact 1 and 2. NL voltages induced by spin injection into Si and creation of a spin current were detected between contact 3 and 4. From comparison between two and four terminals resistances in the sample A at 8 K, the resistance area (RA) product and channel resistivity at low bias (0.05 V) are estimated as about 5 k$\Omega\mu m^2$ and $1.7 \times 10^{-3}$ $\Omega$cm, respectively.

Figure 2 (a) shows the results of NL magnetoresistance (MR) measurements at 8K in Sample A when 0.05 V was applied to the samples (the injection current was about 50 μA). We note that the constant background voltages produced by an electric coupling between electric pads were subtracted from the raw data. NL MR hysteresis appeared at

±18 and ±37 mT [Fig. 2 (a)], where AMR hysteresis [Fig. 2 (b)] was also observed in dummy samples. The clear NL MR signals were observed, whilst any magnetic field dependent signals were not observed in a "local" scheme, where an electric current was injected and an output voltage was measured between two FM electrodes. This indicates that the NL signals were not caused by cross-talk with magnetic field dependent effects like AMR effect, Hall effect, and local MR effect through an unexpected electric pad-to-pad coupling. In addition, $\Delta V$ exhibited apparent exponential decrease with increasing the electrode gap length as shown in an inset of Fig. 2(b). If the observed NL signals are spurious signals, the experimentally observed exponential dependence should not be seen. Hence, we can elucidate that spins were injected into Si and detected at 8 K.

Clear plateaus on the voltage peaks show successful achievement of the anti-parallel magnetization alignment. The spin diffusion length ($\lambda_N$) was estimated to be about 2.25 μm at 8 K from this experimental result [see an inset of Fig. 2(b)] and the following theoretical expression of the output signal intensity [15],

$$V_{non-local} = \beta_i \beta_i' r_N e^{-L/\lambda_N} I_0,$$

where $\beta_i$ and $\beta_i$' are the spin polarization of conducting spins at the contacts 2 and 3, respectively, $r_N$ is spin resistance of Si, $L$ is the gap length between contact 2 and 3, and $I_0$ is the input electric current between contact 1 and 2. The estimated spin diffusion length is comparably larger than that estimated in single- [16] and multi-layer [17] graphene, but much smaller than that in Si-based hot-electron transistors [10]. However, this is not surprising because the impurity concentration of our device is quite high ($1 \times 10^{20}$ cm$^{-3}$) and there should be many spin scattering centers in the channel layer. It should be emphasized that the spin coherent length is larger than 2 μm in spite of such high impurity concentrations. From obtained spin diffusion length and resistivity, the spin resistance of Si at 8 K was calculated to be 38 $\Omega\mu m^2$. Absence of the local MR effect in our samples is a consequence of too high junction resistance (about 5 k$\Omega\mu m^2$) compared to the spin resistance of the Si (38 $\Omega\mu m^2$) [18].

In the case of sample A with a 0.48 μm gap, $\Delta V$ was 0.29 μV when the injection current was set to be 50 μA at 8 K. This experimental observation allows us to calculate the spin polarization of the FM contacts ($\beta_i$ and $\beta_i$') to be 0.02 if we assume that $\beta_i$ and $\beta_i$' are the same. In the case of sample B with a 1.08 μm gap, the spin polarization was estimated to be 0.028. These facts indicate that spin injection efficiency for our samples were not so high, which may depend on the interface condition between the MgO layer and Si interface.

The AC injection current dependence of $\Delta V$ in sample B is shown in Fig. 3. In an inset in Fig. 3, the *I-V* characteristics measured between contact 1 and 2 at 8 K with using a DC electronic current is shown. Very low current below about 0.5 V indicates existence of the tunnel barrier as designed. It is found that $\Delta V$ was proportional to the injection current below 1 mA and becomes saturated above it. Since $\Delta V$ is proportional to spin polarization at contacts in a model above, it can be concluded that $\beta$ is constant below 1 mA. Even after the bias current of 2.5 mA was applied, the observed dependence of the output signals, $\Delta V$, was reproduced, which indicates that an MgO/Si interface was not damaged above 1 mA namely; there was no breakdown or degradation of the MgO barrier.

The temperature dependence of the $\Delta V$ in sample A is shown in Fig. 4, when the bias voltage of 0.25 V was applied at all temperatures. The spin injection signals were observed up to 120 K, which is the highest temperature where band transferred spins are injected in Si to our best knowledge. The signals decreased as the temperature increased. Although mechanism of the reduction is not understood, the increase in the carrier concentration and phonon scattering at elevated temperature are possible origins of it.

In summary, we have succeeded in injecting electron spins into Si through MgO tunnel barrier contact up to 120 K, which is the highest temperature at which band transport spins are injected into Si to our best knowledge. We have estimated the spin diffusion length which is 2.25 μm although an impurity concentration was comparably high ($1\times 10^{20}$ cm$^{-3}$), by achieving anti-parallel magnetization alignments in the spin valves. The results show a feasibility of the doped Si as for a channel material and Fe/MgO as for the spin injector/detector to construct new kinds of spintronics devices.

Acknowledgement

The authors would like to express thanks to H. Nakanishi (AIT) and K. Namba (TDK) for supporting this research. The authors are grateful to T. Yamane and N. Mitoma (Osaka University) for valuable discussions and the experimental support, and to Y. Ishida, M. Kubota, S. Tsuchida, and K. Yanagiuchi (TDK) for experimental support in sample analysis and fruitful discussions. One of the authors (Y.S.) wishes to acknowledge Dr. W. Van Roy of IMEC for helpful discussions.

Figure captions

Fig.1 (a) A schematic of depiction for the Si spin valve with four terminals. An external magnetic field is applied along +*y* and -*y* directions. A spin polarized current is injected from contact 2 to contact 1, and the spin injection signals are detected at contact 3 and contact 4 as an output voltage. (b) The inset shows a optical image of the actual device.

Fig.2 (a) Result of a non-local MR measurements at 8 K in sample A with the gap length of 0.48 $\mu$ m. The devices were biased at 0.05 V. The shown data are after a removal of an offset voltage and normalization at 50 µA. The measurements were repeated 4 times and averaged in order to obtain clear data. (b) AMR hysteresis curves measured in a dummy sample between contact 1 and 2 (blue line), and between contact 3 and 4 (pink line). The inset shows the contact gap dependence of the output voltage in sample A and B.

Fig.3 The injection current dependence of the spin-accumulation-induced voltage at 8 K in sample B with the 1.08 µm gap. The inset shows an *I-V* characteristic in the DC electric current injection circuit at 8 K.

Fig.4 Temperature dependences of the spin-accumulation-induced voltage measured in sample A blow 150 K.

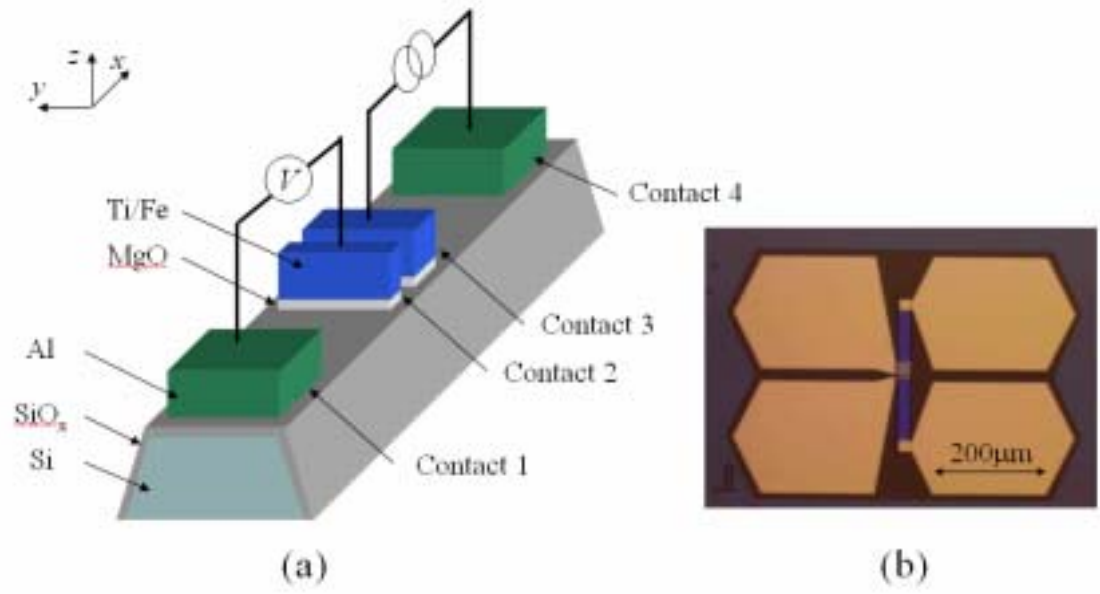

Fig.1

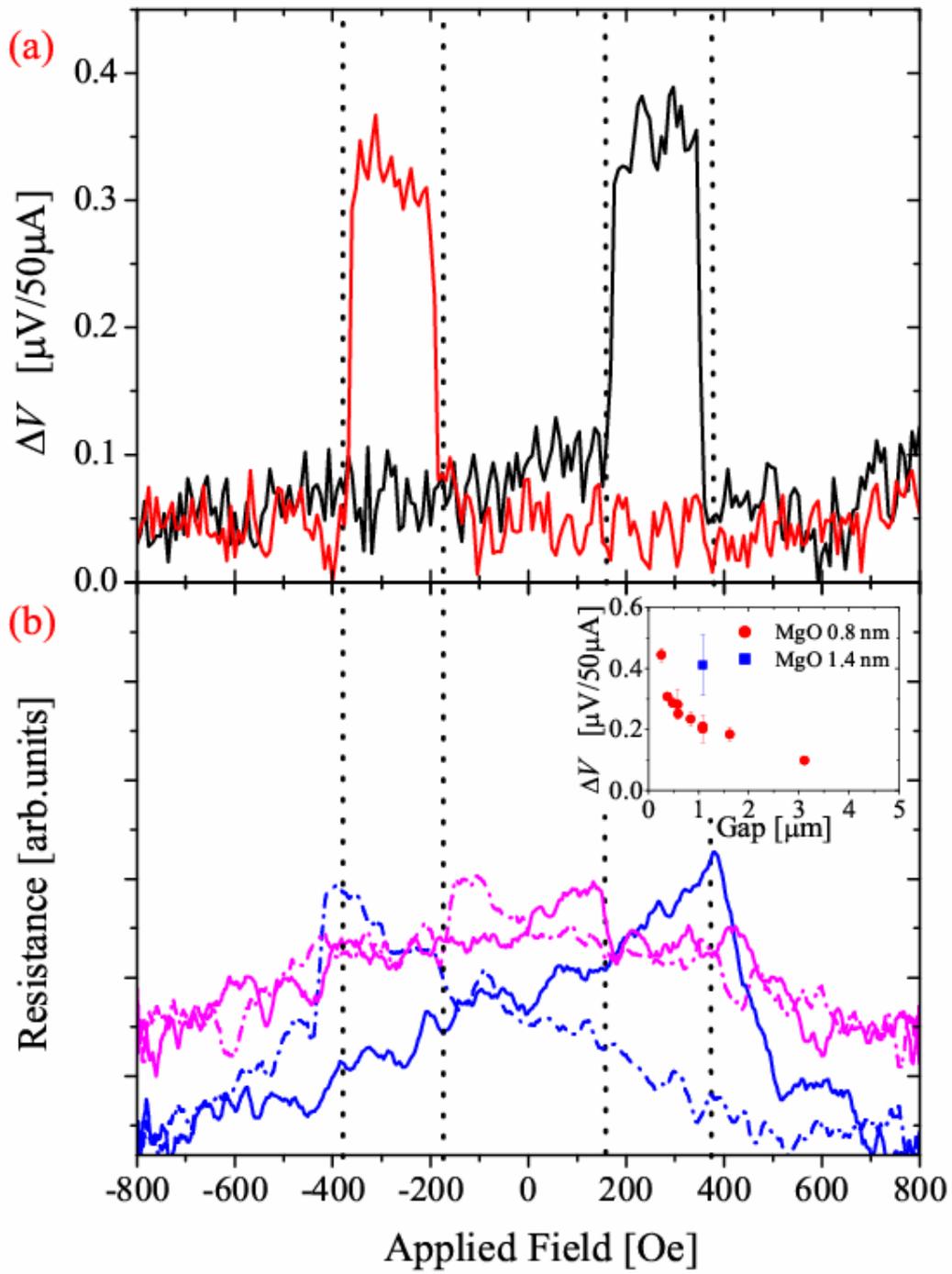

Fig.2

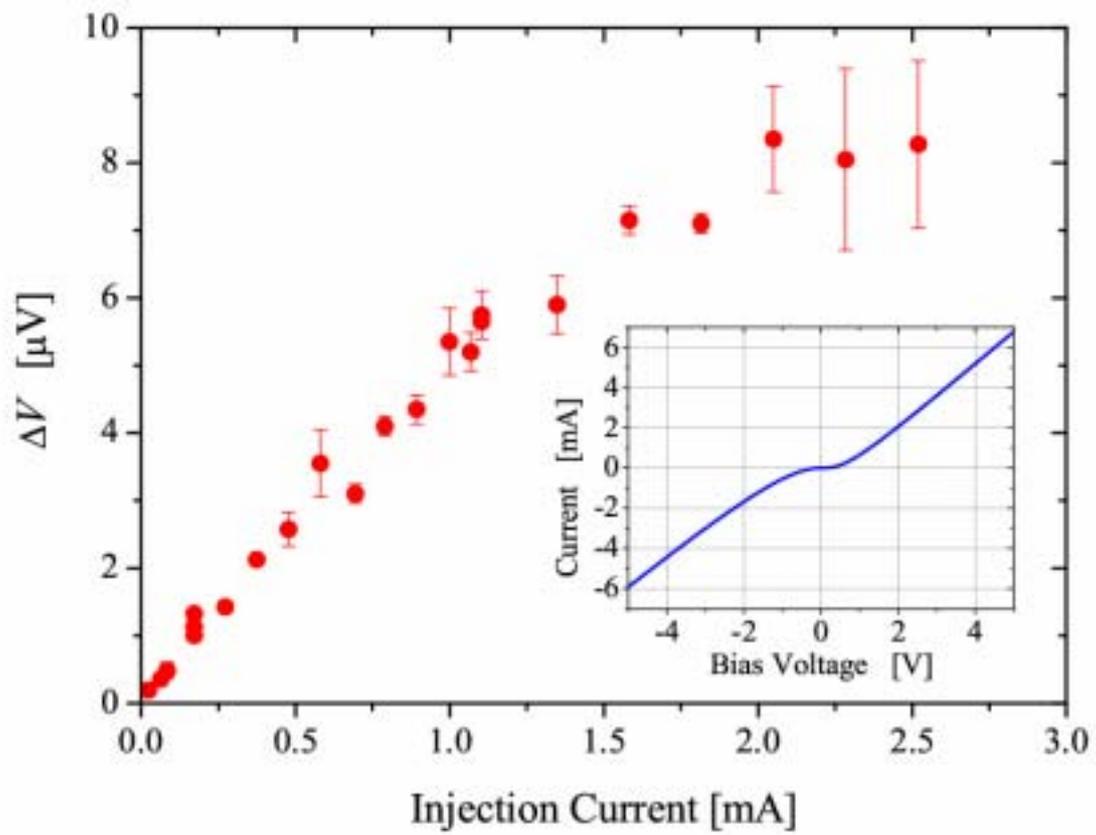

Fig.3

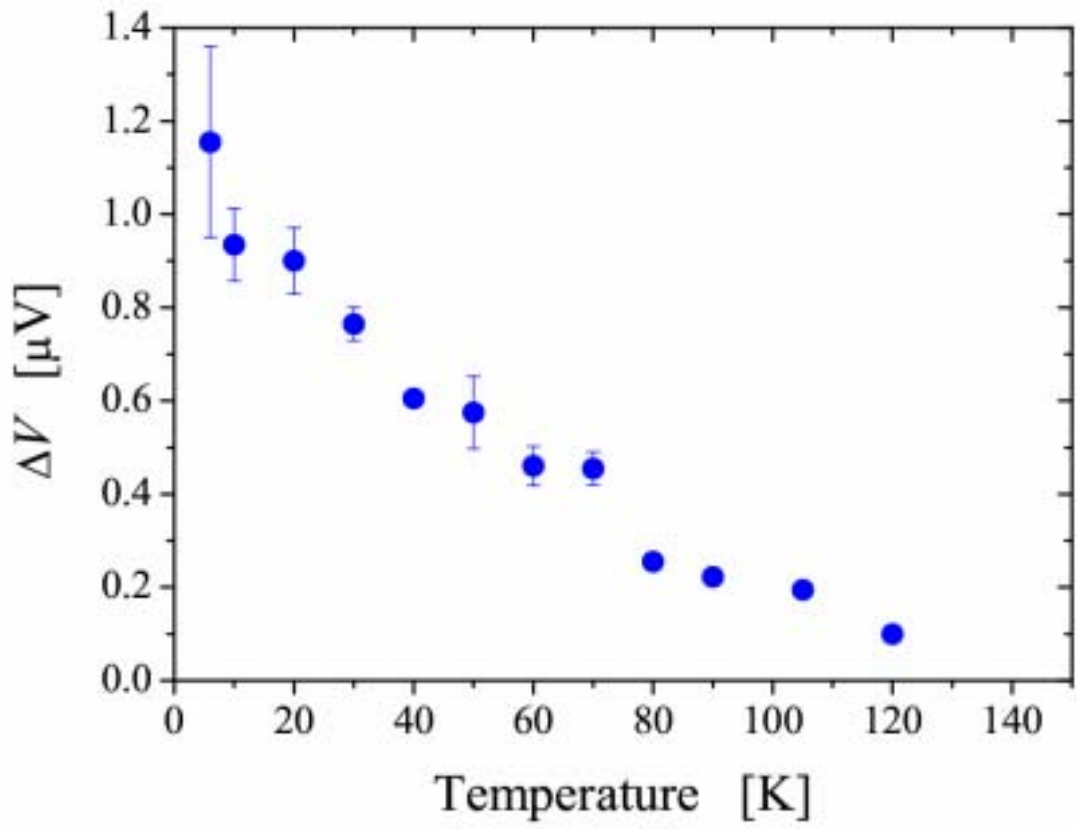

Fig.4